\documentclass[aps,pra,reprint,superscriptaddress,showpacs,showkeys]{revtex4-2}
\usepackage{graphicx}
\usepackage{siunitx}

\begin{document}

\title{Preparation of ultra-cold atomic-ensemble arrays using
time-multiplexed optical tweezers}

\author{Katja Gosar}
\email[]{katja.gosar@ijs.si}
\affiliation{Jo\v{z}ef Stefan Institute, Jamova 39, SI-1000 Ljubljana, Slovenia}
\affiliation{Faculty of Mathematics and Physics, University of Ljubljana, Jadranska 19, SI-1000 Ljubljana, Slovenia}

\author{Vesna Pirc Jev\v{s}enak}
\affiliation{Jo\v{z}ef Stefan Institute, Jamova 39, SI-1000 Ljubljana, Slovenia}
\affiliation{Faculty of Mathematics and Physics, University of Ljubljana, Jadranska 19, SI-1000 Ljubljana, Slovenia}

\author{Tadej Me\v{z}nar\v{s}i\v{c}}
\affiliation{Jo\v{z}ef Stefan Institute, Jamova 39, SI-1000 Ljubljana, Slovenia}
\affiliation{Faculty of Mathematics and Physics, University of Ljubljana, Jadranska 19, SI-1000 Ljubljana, Slovenia}

\author{Du\v{s}an Babi\v{c}}
\affiliation{Aresis Ltd, Ulica Franca Mlakarja 1a, SI-1000 Ljubljana, Slovenia}

\author{Igor Poberaj}
\affiliation{Aresis Ltd, Ulica Franca Mlakarja 1a, SI-1000 Ljubljana, Slovenia}

\author{Erik Zupani\v{c}}
\affiliation{Jo\v{z}ef Stefan Institute, Jamova 39, SI-1000 Ljubljana, Slovenia}

\author{Peter Jegli\v{c}}
\email[]{peter.jeglic@ijs.si}
\affiliation{Jo\v{z}ef Stefan Institute, Jamova 39, SI-1000 Ljubljana, Slovenia}

\date{\today}

\begin{abstract}
We use optical tweezers based on time-multiplexed acousto-optic deflectors to trap ultra-cold cesium atoms in one-dimensional arrays of atomic ensembles. For temperatures between \SI{2.5}{\micro K} and \SI{50}{nK} we study the maximal time between optical tweezer pulses that retains the number of atoms in a single trap. This time provides an estimate on the maximal number of sites in an array of time-multiplexed optical tweezers.  We demonstrate evaporative cooling of atoms in arrays of up to 25 optical tweezer traps and the preparation of atoms in a box potential. Additionally, we demonstrate three different protocols for the preparation of atomic-ensemble arrays by transfer from an expanding ultra-cold atomic cloud. These result in the preparation of arrays of up to 74 atomic ensembles consisting of $\sim$100 atoms on average.
\end{abstract}


\maketitle

\section{Introduction}

Optical tweezers is a common name for steerable optical dipole traps.
In cold atom experiments optical tweezers are used for painting arbitrary static and dynamic potentials. This includes creating arrays of optical tweezers at arbitrary positions \cite{wang2020preparation, endres2016atom, bernien2017probing, pu2018experimental}, moving of such optical traps \cite{carpentier2008laser, beugnon2007two, gustavson2001transport, rakonjac2012laser} and painting box, ring \cite{bell2016bose} and other shapes of optical potentials. The positional control of the traps can be achieved with several different methods: using acousto-optic deflectors (AODs), spatial-light modulators (SLMs) \cite{boyer2006dynamic, muldoon2012control} or  digital micromirror devices (DMDs) \cite{wang2020preparation, gauthier2016direct, stuart2018single}. In this work we focus on optical tweezers based on AODs. 

Deflection of a laser beam on an AOD is controlled by the frequency of the drive signal. When creating multiple optical traps or painting an arbitrary potential, AODs can be used in two different ways (i) by using continuous multi-frequency driving, where the AODs are simultaneously driven with all frequencies corresponding to trap positions \cite{shin2004atom, endres2016atom, trypogeorgos2013precise}, or (ii) by using the time-multiplexed approach, where only a single frequency is applied at a time, but the frequencies are rapidly switching to create a time-averaged optical potential \cite{schnelle2008versatile, bell2016bose, chisholm2018three, deb2014optical}. The use of the latter approach is studied in this article.

In our experiments the optical traps are created at the intersection of optical tweezers with a dimple beam. In principle, a single-axis AOD would allow the control of the position of the tweezers along the dimple, however that would call for perfect alignment of the AOD axis with the dimple beam \cite{endres2016atom}. We use a two-axis AOD which allows us to control the position of optical tweezers in two dimensions. In the case of two-axial AODs the multi-frequency approach creates ``ghost'' traps, which lie outside the dimple beam, causing a loss of optical power into the unused traps \cite{roberts2014steerable}. Because of the ghost traps multi-frequency driving cannot produce an arbitrary potential in two-dimensions, but with time-multiplexing an arbitrary potential can be painted \cite{onofrio2000surface, henderson2009experimental}. For potentials with a suitable symmetry, a combination of the multi-frequency and time-multiplexed approach can be implemented \cite{zimmermann2011high}.

\begin{figure}[b]
\includegraphics[width=1.0\linewidth]{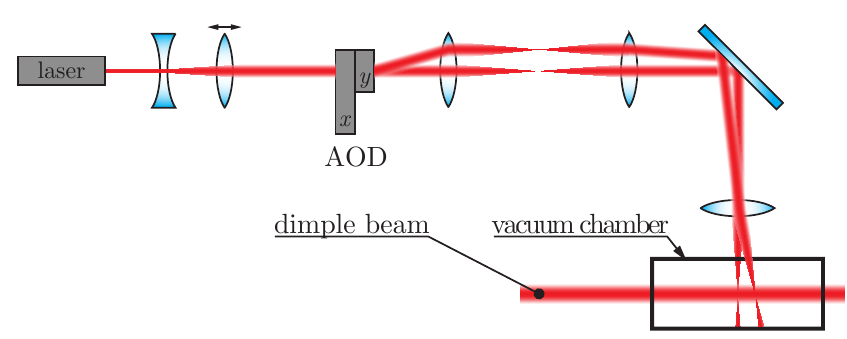}
\caption{Experimental setup schematically showing the optical tweezers setup. The position of the optical tweezers focal plane is controlled by the position of a lens before the AODs. The two-dimensional position of the optical tweezers is controlled by the AODs. The optical tweezers cross the horizontal dimple beam inside the experimental vacuum chamber. The guide beam (not shown) is perpendicular to the page.  
}
\label{fig1}
\end{figure}

\begin{figure*}
\includegraphics[scale=1]{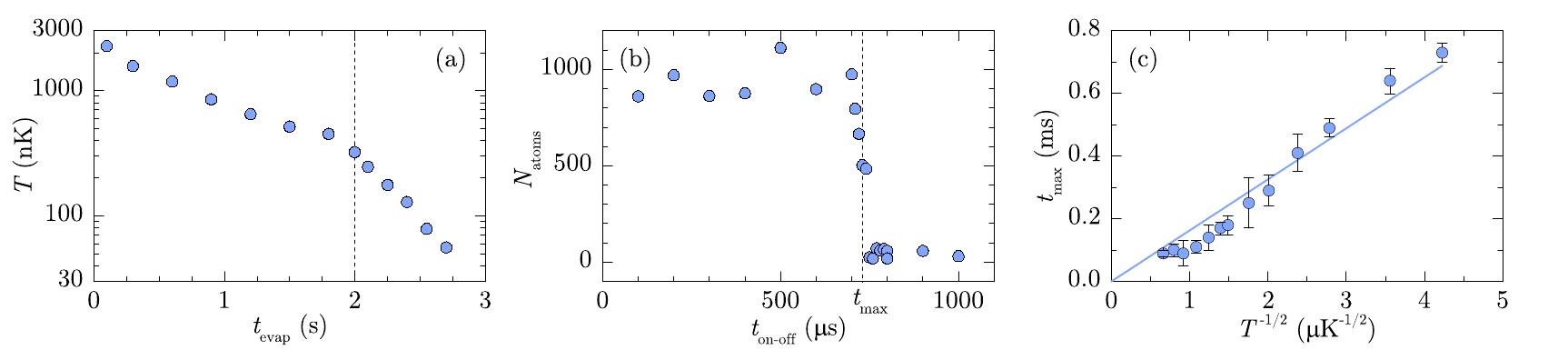}
\caption{Characterisation of multiplexing. (a) Temperature of the atomic cloud as a function of the evaporation time in a single optical tweezer trap. The vertical dashed line marks the start of the magnetic-field gradient rampdown. (b) An example of a $t_{\mathrm{max}}$ measurement ($t_{\mathrm{evap}} =$ \SI{2700}{ms}), showing the number of atoms retained in the trap for different $t_{\mathrm{on-off}}$. $t_{\mathrm{max}}$ is determined as the time, for which the number of retained atoms rapidly drops. (c) Maximal time $t_{\mathrm{max}}$ between optical tweezer pulses that maintains the number of atoms in the trap, as a function of the temperature and the linear fit to the measured data. 
}
\label{fig2}
\end{figure*}

Preparation of ultra-cold atomic-ensemble arrays is motivated by the possibility of using them as quantum simulators using Rydberg atoms. Previously, this type of quantum simulation was performed using arrays of single atoms \cite{bernien2017probing, endres2016atom, wu2021review, ebadi2021quantum}, but it is also possible to use collective excitations in ensembles of a few hundred atoms  \cite{ebert2015coherence, wang2020preparation, spong2021collectively}. With ensembles the preparation and detection of Rydberg states can be significantly faster than with single atoms \cite{xu2021fast}.

We demonstrate four different methods of preparing a one-dimensional array of atomic ensembles in time-multiplexed optical tweezers. The first method is based on loading the traps directly from the large dipole trap and performing simultaneous evaporative cooling with them. In the other three methods the first step is evaporative cooling of an atomic cloud in the dimple trap. It is followed by different forms of expansion of the atomic cloud along the dimple beam and trapping into an array of optical tweezers.

In addition to optical tweezer arrays, optical tweezers are also used for the preparation of arbitrary time-averaged and dynamic potentials. These can be used to perform experiments with solitons \cite{ismailov2021confinement, carpentier2008laser} or to prepare condensates in different shaped traps \cite{navon2021quantum}. In this work we present examples of painting a box potential and a one-dimensional harmonic potential. As an example of a dynamic time-averaged potential, we present splitting of atomic ensembles.

\section{Experiment}

We prepare cold $^{133}$Cs atoms by laser cooling with the standard procedure described in detail in Ref. \onlinecite{meznarsic2019cesium}. All experiments presented in this article start with \SI{3e6}{} atoms in a large dipole trap, cooled to \SI{2}{\micro K}. 
For evaporative cooling we use a small dipole trap inside the large dipole trap. It is created by crossing two laser beams that we call the dimple beam and the guide beam. The dimple beam has a $1/\mathrm{e}^2$ radius of  \SI{32}{\micro m} and the guide \SI{60}{\micro m}. Instead of the guide beam, we can use optical tweezers.
Our experimental setup setup with commercial optical tweezers (Aresis, CALM 1064) and the dimple beam is illustrated in Fig. \ref{fig1}. The radius of the optical tweezers beam waist is $w_0 \approx \SI{2.1}{\micro m}$, and the Rayleigh length is $z_0 \approx \SI{13.0}{\micro m}$. To create a trap that holds the atoms against gravity the tweezers have to be combined with another beam. In all our experiments with optical tweezers the traps are created by crossing a tweezer beam with the dimple beam. The position of the optical tweezers is optimized so that they are centered with the dimple beam. This is important because misalignment of the two beams leads to leakage of atoms from the trap \cite{roberts2014steerable}. The accessible area of our optical tweezers is \SI{1.7}{mm} $\times$ \SI{1.7}{mm} and we calibrate the AOD drive amplitude so that the intensity of the tweezers is independent of the trap position. Our optical tweezers setup features an additional lens that can move the tweezers' beamwaist, as shown in Fig. \ref{fig1}. For experiments in section A and B the crossing of optical tweezers with the dimple beam is approximately \SI{14}{\micro m} from the tweezers' beamwaist, where the radius of the optical tweezers beam is \SI{3.1}{\micro m}. For experiments in section C, the tweezers are used at the beamwaist.

\subsection{Temperature dependence of the maximal trap switching time}
\label{section II A}

We prepare an array of optical tweezers traps by time multiplexing with the switching frequency $\nu =$ \SI{100}{kHz}, which is the upper limit for our optical tweezers. This means that each trap is active for only \SI{10}{\micro s} in each cycle and then it is off for $t_{\mathrm{off}} = \frac{N-1}{\nu}$, where $N$ is the number of traps in the array. That is why we first characterise the dependence of $t_{\mathrm{off}}$ on the temperature. This gives us an estimate on the number of traps in an array that would efficiently trap an atomic ensemble at a given temperature.

We start by transferring atoms from the large dipole trap into a single optical tweezer trap in which we then evaporatively cool the atoms. First, the power of the tweezer and dimple beam is linearly ramped up for \SI{1500}{ms}. Then, we hold the power for \SI{400}{ms} to let the atoms thermalize. Afterwards, we turn off the large dipole trap and start the evaporation cooling. The power of the beams is ramped down exponentially ($\tau_{\mathrm{dimple}} =$ \SI{0.625}{s}, $\tau_{\mathrm{tweezers}} =$ \SI{1}{s}) for the evaporation time $t_{\mathrm{evap}}$, asymptotically approaching 6.8\% (dimple) and 5\% (tweezers) of the maximum power. After \SI{2}{s} of evaporation, we start with an additional exponential rampdown of the magnetic field gradient, which contributes to more efficient evaporative cooling \cite{meznarsic2019cesium}. We measure the temperature of the atoms as a function of the evaporation time $t_{\mathrm{evap}}$ (Fig. \ref{fig2}(a)). After \SI{3}{s} of evaporation, the atoms form a Bose-Einstein condensate (BEC), which can be seen from the standard time-of-flight measurement.
 
To estimate the maximal possible number of traps in an array, we measure the maximal time between optical tweezer pulses that retains the number of trapped atoms and its dependence on the temperature of the atoms. During the evaporation, the optical trap is switched on and off with \SI{100}{kHz}; the trap is on for \SI{10}{\micro s} and then off for \SI{10}{\micro s}. After a given evaporation time $t_{\mathrm{evap}}$, we stop the evaporation by keeping the trap powers constant. The tweezer trap is then switched on and off with pulses of length $t_{\mathrm{on-off}}$ and a 50\% duty cycle. 
This way the time-averaged power of the trap is independent of the $t_{\mathrm{on-off}}$ when using the same amplitude. We measure the number of atoms that are retained in a pulsing trap after \SI{100}{ms} as a function of $t_{\mathrm{on-off}}$. An example of this measurement is shown in Fig \ref{fig2}(b). The number of atoms is constant for sufficiently short $t_{\mathrm{on-off}}$ and  it drops off to zero at time $t_{\mathrm{max}}$. The colder the atom cloud is, the longer the  $t_{\mathrm{max}}$.

$t_{\mathrm{max}}$ is determined by the velocity of the atoms and the size of the optical tweezer trap. For the atoms to be recaptured with the next pulse, the average distance $d$ an atom travels in the time $t_{\mathrm{on-off}}$ has to be smaller than the effective radius of the trap $w$ \cite{tuchendler2008energy, he2011extending}. We calculate $d$ as
\begin{equation} \label{d}
d = \sqrt{\frac{2k_B T}{m}} t_{\mathrm{on-off}},
\end{equation}
where $\sqrt{2 k_B T/m}$ is the average (rms) velocity of an atom of an ideal gas in two dimensions at temperature $T$. $k_B$ is the Boltzmann constant and $m$ is the mass of a cesium atom. We use the two dimensional average velocity because atoms escape the optical tweezers only in the radial direction.
Since the dimple beam is approximately 10-times wider then the tweezers we can neglect its effect on this process.  From Eq. (\ref{d}) we can conclude  that $t_{\mathrm{max}}$ is proportional to $T^{-1/2}$ as
\begin{equation} \label{tmax}
t_{\mathrm{max}} = \sqrt{\frac{m}{2k_B T}} w.
\end{equation} 
Here we assume that the effective trap radius $w$ is constant.
We plot $t_{\mathrm{max}}$ as a function of $T^{-1/2}$ in Fig. \ref{fig2}(c). The linear fit to the data gives us an estimate of the effective radius of the optical tweezers $w = $ \SI{1.8(1)}{\micro m}.
The deviation of the measured $t_{\mathrm{max}}$ from the model could be explained by the dependence on the trapping width on the optical power, the effect of the dimple beam and the magnetic field gradient, which additionally deforms the trap geometry.

\begin{figure}[h!]
\centering
\includegraphics[width=\linewidth, trim = 0 25 0 0, clip]{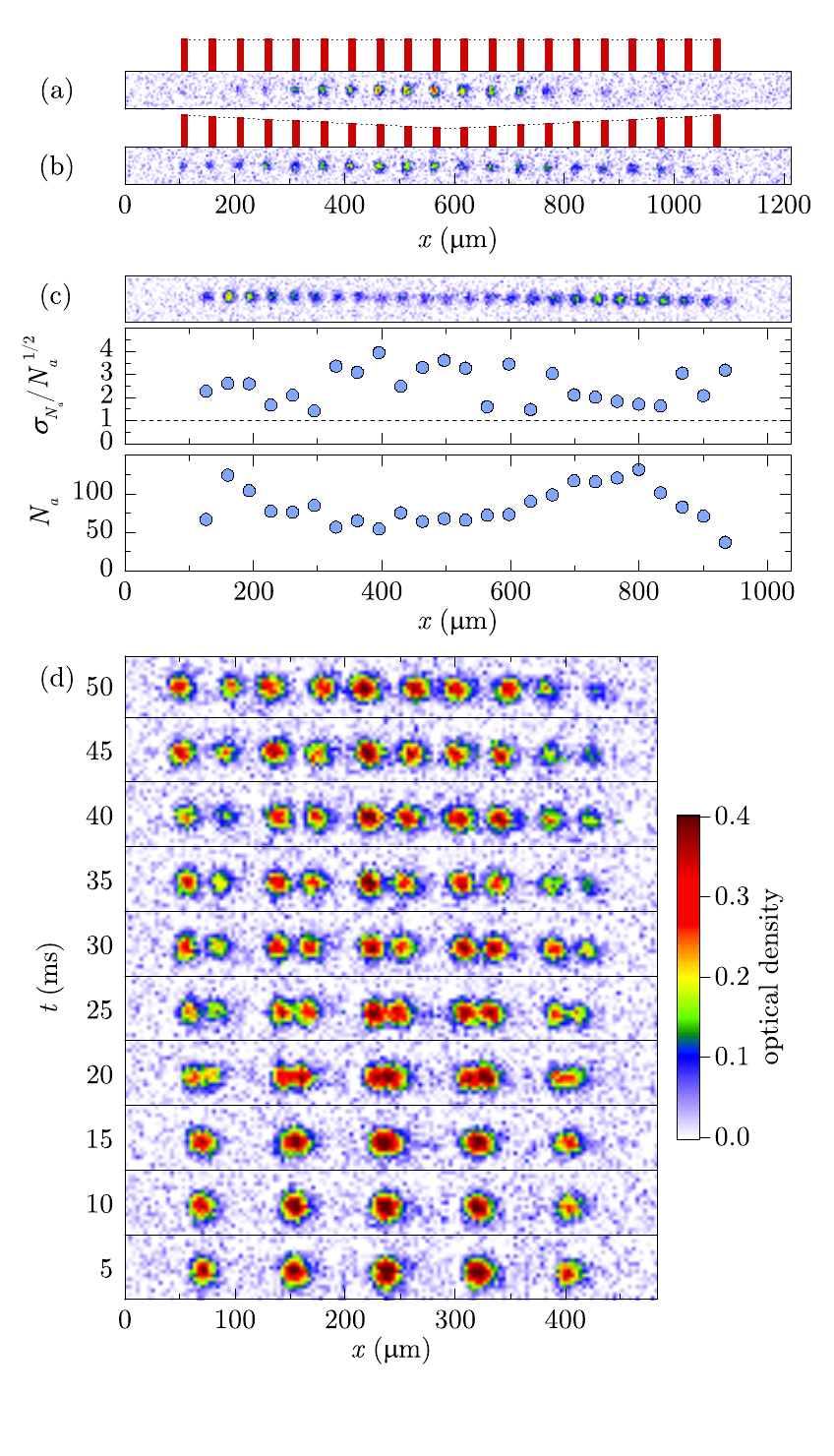}
\caption{Arrays loaded directly from a dipole trap.  (a) An absorption image of atoms in an array of 20 tweezers of equal intensity. (b) Atoms in an array of 20 sites with a V-shaped profile of tweezer intensities. All other parameters are the same as in image (a). The red bars illustrate the intensities of the optical tweezers. (c) Absorption image of a 25 site array of atomic ensembles in optical tweezers, loaded directly from the large dipole trap. The image is an average of 10 realisations. Below, we plot $\sigma_{N_a}/\sqrt{N_a}$ and $N_a$, where $\sigma_{N_a}$ is the variance of the number of atoms within each trap determined from 10 measurements and $N_a$ is the average number of atoms in each trap. The dashed line marks $\sigma_{N_a}/\sqrt{N_a} = 1$ which represents the Poissonian distribution of the number of atoms in each trap. (d) Splitting of 5 atomic ensembles into 10. Absorption images taken at \SI{10}{ms} intervals of the \SI{100}{ms} splitting. All images use the same colorbar, except (c) where the same colors correspond to optical densities from 0 to 0.2. The optical tweezers are time-multiplexed with a frequency of \SI{100}{kHz}. This means that each trap is active for \SI{10}{\micro s} and its duty cycle is $1/N$, where $N$ is the number of traps forming the array.}
\label{fig3}
\end{figure}

\subsection{Loading of an array and a box potential directly from the dipole trap}
\label{section II B}

\begin{figure*}
\includegraphics[scale=1]{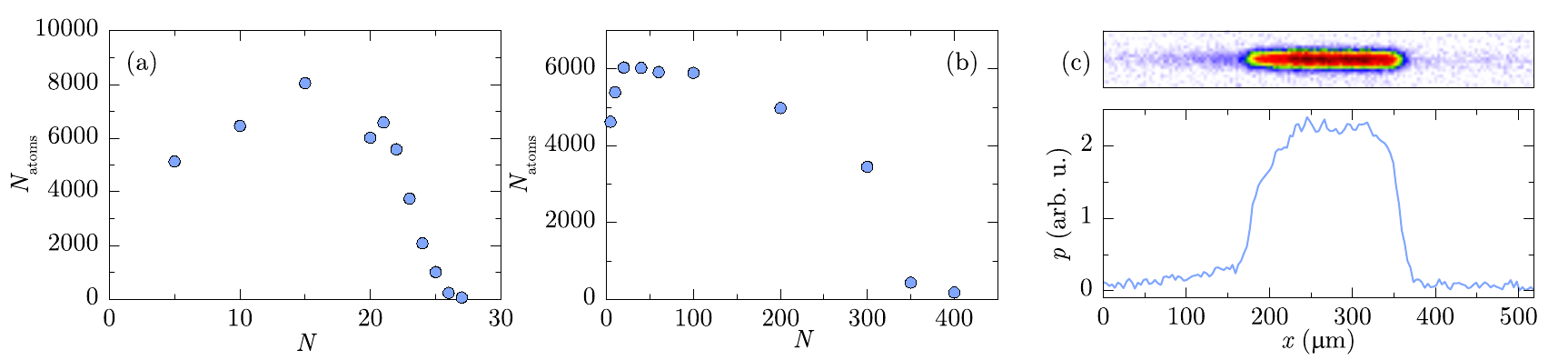}
\caption{Characterisation of the number of traps forming the potential for preparation of cold atoms in a line box-shaped potential. (a) The number of atoms after \SI{3000}{ms} of evaporation in a box-shaped potential as a function of the number of traps $N$ forming the box. (b) The number of atoms in a box potential after \SI{2800}{ms} of evaporation in an array of 5 tweezers and  additional \SI{400}{ms} of evaporation in a box potential formed by $N$ tweezer traps. (c) Absorption image (colorbar in Fig. \ref{fig3}) and the linear density profile $p$ of atoms in a box-shaped potential, prepared with a transfer from 5 to 40 sites. Average of 10 measurements.}
\label{fig4}
\end{figure*}

\begin{figure}[h!]
\includegraphics[width = \linewidth, trim = 0 25 0 0, clip]{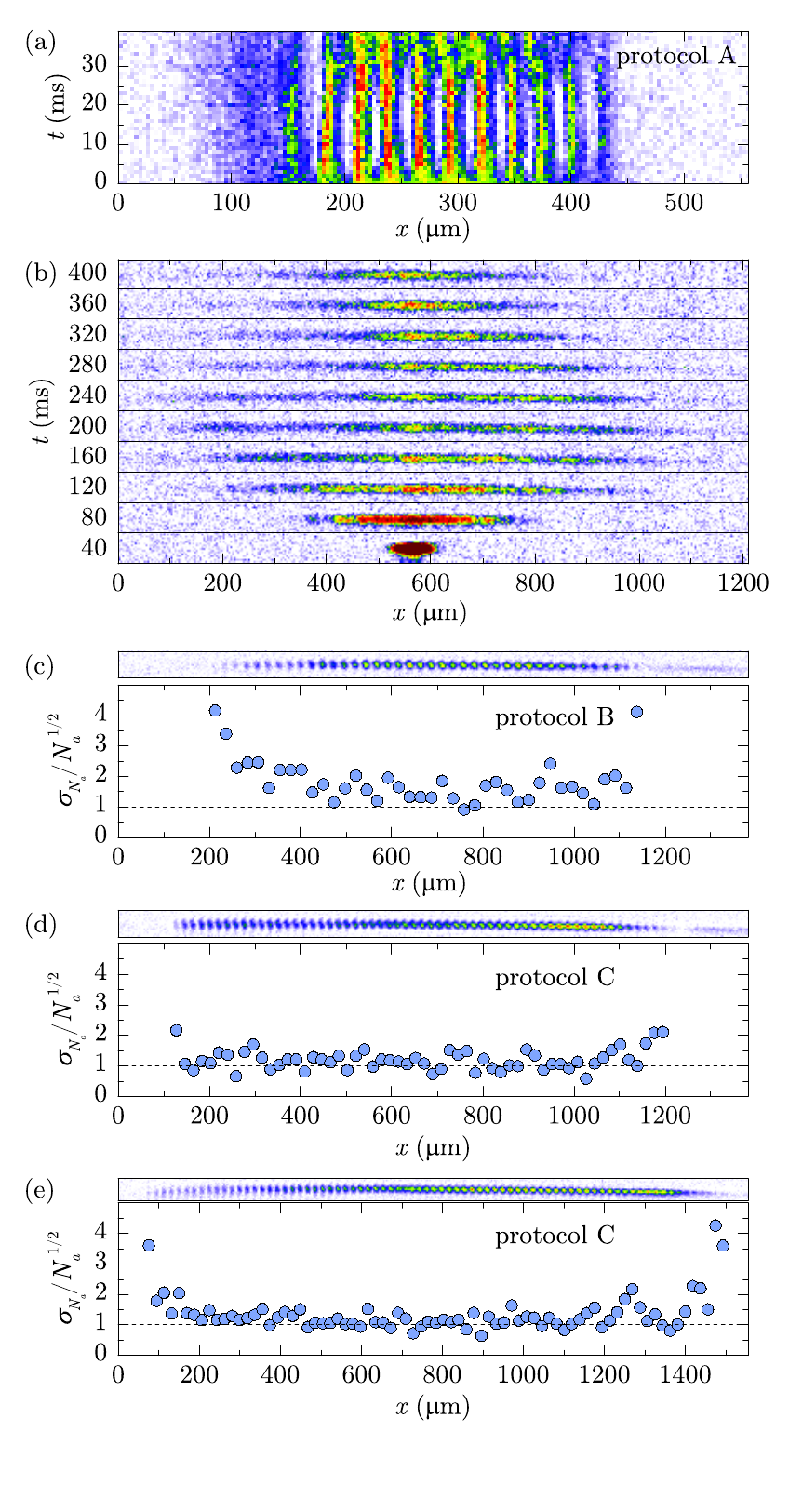}
\caption{Arrays loaded from a dimple trap using one-dimensional time-of-flight or expansion in a harmonic potential. Protocol A: loading the tweezers from an expanding cloud. (a) shows the linear density of the atomic ensembles in an array of 10 optical tweezers as a function of time after the optical tweezers are turned on. A breathing mode oscillation can be seen. Protocol B: expansion of the atomic cloud in a painted harmonic potential and switching to an array of optical tweezers after half the breathing oscillation. (b) shows the oscillation of the atomic cloud in the painted harmonic potential and (c) the resulting array (optical density image and the normalized variance $\sigma_{N_a}/\sqrt{N_a}$). Protocol C: ramping the optical tweezer array power and position with the expansion of the atomic cloud. (d) and (e) show the results of protocol C with 60 and 80 traps (74 filled), respectively. (c), (d) and (e) show an average of 10 realisations of the experiment and the variance is determined from the same 10 measurements. The colorbar is the same as in Fig. \ref{fig3}.}
\label{fig5}
\end{figure}

The simplest method of preparing an array of atomic ensembles with optical tweezers is simultaneous evaporation in the optical tweezers.  The protocol for evaporative cooling is identical to the one presented in Section \ref{section II A}, except for the larger number of optical tweezer traps. We observe that the distribution of atoms in the array reflects the atom density profile of the dipole trap. We were able to achieve more uniform filling by adjusting the power in the tweezers to a V-profile; traps on the edge of the array have more power than those in the middle. Examples of arrays created with a uniform strength profile and a V-profile are shown are Fig. \ref{fig3}(a) and \ref{fig3}(b). The largest array we are able to prepare by trapping directly from the large dipole trap has 25 atomic ensembles.  As shown in Fig. \ref{fig3}(c), the variance of the number of atoms per trap is relatively large.
This method is similar to experiments in Ref. \onlinecite{deb2014optical}, where evaporative cooling to BEC in four time-multiplexed optical traps was achieved. In contrast to our method, the atomic clouds for evaporation were prepared with splitting. 

Splitting of the atomic ensembles can also be done after the evaporative cooling, as a way to achieve a higher number of traps in an array. The atoms are cooled in a smaller array and then the number of array sites is doubled by splitting the atomic ensembles \cite{roberts2014steerable}. Fig. \ref{fig3}(d) shows the splitting of an array of 5 traps into 10. The splitting starts after \SI{2800}{ms} of evaporation and is finished in \SI{100}{ms}. The tweezers' trajectories are linear. The challenge with splitting is ensuring even distribution of atoms in the split traps, which can be achieved by precise tuning of the traps' position and intensity. Additionally, the splitting could be improved by using a minimum jerk cost trajectory of the tweezers, as it was done in Ref. \onlinecite{roberts2014steerable}.

If the distance between the traps in the array is smaller than the width of a trap, they form a box potential. We measured the number of atoms in an array of $N$ traps that span \SI{185}{\micro m}, after \SI{3000}{ms} of evaporation. The measured number of the atoms as a function of the number of traps is shown in Fig. \ref{fig4}. For $N < 15$ the number of atoms increases with the number of traps. This is because the traps are separate and an additional trap can trap additional atoms. For $N > 20$, we see a drop in the number of atoms. Here the traps still do not form a box potential, but the number of trapped atoms lowers due to larger $t_\mathrm{off}$.

If we wanted to prepare a larger box potential, we would need a larger number of traps, but that also leads to higher $t_\mathrm{off}$ that has an adverse effect on trapping. However, the lower the temperature of the atoms, the larger number of traps we can use. This gives the idea of cooling the atoms in an array with a lower number of traps and then, when the atoms are sufficiently cooled, switching to a larger number of traps, that cover the same area, forming a box potential. For example, we start by evaporatively cooling atoms in an array of 5 optical tweezer traps. After \SI{2800}{ms} of evaporation the temperature is \SI{48}{nK}. We then switch to an array of $N$ tweezers, and keep the power of the beams constant for \SI{400}{ms} to let the atoms cover the \SI{185}{\micro m} box potential. Fig. \ref{fig4}(c) shows an example profile of a filled box potential. Fig. \ref{fig4}(b) shows the final number of atoms as a function of $N$. For small $N$ we see the same effect as with loading directly into a box potential: the traps are separated and the number of atoms increases with the number of traps. For more than 100 traps, the number of atoms starts decreasing. This is the regime were the trap separation is the same as the trap radius and the traps truly form a box potential. The number of atoms decreases with $N$ due to the larger $t_{\mathrm{off}}$. With more than 400 traps no atoms are trapped in the box potential. This is due to atom losses at the edges of the box potential, if the traps at the edges are not repeated with a high enough frequency.

\subsection{Loading from the dimple trap}
\label{section II C}

We see that for creating larger atomic-ensemble arrays lower temperatures of atoms are needed. This is why we developed alternative methods of preparing atomic-ensemble arrays. In these methods, the atoms are first cooled in the crossed dimple and guide trap and then transferred into an array of optical tweezers. However, the transfer can lead to unwanted oscillations of the atomic ensembles inside the optical traps. We present three different protocols that tackle this problem in different ways.

The first protocol starts after \SI{4500}{ms} of evaporative cooling (atoms in BEC), when the guide beam is slowly turned off, which allows the atomic cloud to expand along the dimple beam. The power of the guide beam is decreased as a hyperbolic tangens from the initial power to zero in \SI{1500}{ms}. This ensures slower expansion of the atomic cloud as opposed to a sudden switching off of the guide beam. We allow the cloud to expand for additional \SI{100}{ms} and then turn on an array of 10 optical tweezer traps that span \SI{215}{\micro m}. We will refer to this method as protocol A. The resulting array of 10 atomic ensembles is shown in Fig. \ref{fig5}(a). In the image, 11 bunches of atoms can be seen. The 10 bunches on the right side correspond to atoms trapped by the optical tweezers and the remaining one in the left-most position is the untrapped atoms.

The drawback of protocol A is that the filling of the tweezers is intrinsically non-uniform, because it is determined by the distribution of atom velocities that is mapped into their positions by the expansion. This also means that the atoms, that we attempt to trap at the edges of the array have a higher velocity. This additionally decreases the number of trapped atoms and induces oscillations of atoms inside the traps. We observed that the atomic clouds oscillate inside all the tweezer traps. This is shown in Fig. \ref{fig5}(a), where we plot the linear density of the atoms as a function of time.

We aim to reduce the motion of atoms due to expansion with protocol B. Here, we instantly turn off the guide beam after \SI{3000}{ms} of evaporation, and then paint a harmonic potential with optical tweezers. The harmonic potential is painted in 1000 points on a span of \SI{900}{\micro m} along the dimple beam and it is centred around the position of the crossed dimple trap. The breathing oscillation of the atomic cloud in the painted harmonic potential is shown in fig. \ref{fig5}(b). The frequency of the oscillation is \SI{2.8}{Hz}. After \SI{180}{ms}, which is a half of the oscillation, the atoms are at the maximal amplitude and have, in principle, zero velocity (if the center of the harmonic potential is perfectly aligned with the initial position of the atoms). This is when we stop painting the harmonic potential and instead paint an array of 40 optical tweezers to capture the atoms. The strength of the tweezers is ramped for \SI{100}{ms}, after which we image the atomic-ensemble array. Ramping of the optical tweezer power results in less oscillations in the tweezers compared to a sudden switch to traps at full power. The result is shown in Fig. \ref{fig5}(c). All 40 traps are occupied, but the occupation of the edge traps varies more than of the middle traps, where it even reaches the Poissonian limit $\sigma_{N_a} = \sqrt{N_a}$.

In protocol C, we instantly turn off the guide beam after \SI{3000}{ms} of evaporation (the temperature of the atomic cloud is $\sim$\SI{70}{nK}) and start the \SI{200}{ms} ramp up of the optical tweezer power. We start with all the traps in the center and then, during the ramp-up, linearly expand the array to the final span of \SI{1100}{\micro m}. This is approximately the same rate as the expansion of the atomic cloud in absence of tweezer traps. This way, we follow the motion of the atoms and therefore avoid exciting oscillations in the trap.
An example of an array of 59 atomic ensembles prepared by this protocol is shown in Fig. \ref{fig5}(d). The image is taken at the end of the expansion. We were able to achieve larger arrays with shorter evaporation times. Because of the higher temperature, the rate of expansion is higher, allowing for a larger array. For example, Fig. \ref{fig5}(e) shows an array of 74 filled traps, out of 80 optical tweezers, where the time of evaporation was shortened to \SI{2900}{ms}, corresponding to $\sim$\SI{100}{nK} at the end of evaporation). In this case, the number of filled traps in the array is not limited by temperature but by the rate of expansion, resulting in a larger number of filled traps at the higher temperature. Similarly to protocol B, the variance is larger at the edges of the array and is about $\sigma_{N_a}/ \sqrt{N_a} = 1$ in the middle of the array.

\section{Discussion and Conclusions}

Comparing the $t_{\mathrm{off}} = \frac{N-1}{\nu}$ to the measured $t_{\mathrm{max}}$ at a given temperature gives us an estimate for the maximal number $N$ of traps in an array that would still efficiently trap atoms at that temperature. The observed maximal number of traps is higher. This can by understood through the effect of neighbouring traps; if an atom escapes from a certain trap it is not necessarily ``lost'' as it can be captured by a neighbouring trap.
We note that the temperature of the atoms is not the only limiting factor on the number of traps. The size of the array is also limited by the available trap power and the maximal possible switching frequency. For very large arrays the available optical power would need to be increased to ensure sufficient effective trap depths. The trap frequency sets a constraint on the number of traps in array at a given switching frequency. This is due to parametric heating that occurs when the frequency of the multiplexing seen by the atoms is close to the trap frequency \cite{chisholm2018three}. In our case, the trap frequencies used in Section \ref{section II A} are at most \SI{215}{Hz} at the start of evaporation and \SI{100}{Hz} after \SI{3}{s} of evaporation, which is much lower that the frequencies of on-off switching at which we observed the loss of atoms.

The arrays presented in this work have a high variation in the number of atoms per trap inside an array. In Section \ref{section II B} we demonstrate that the homogeneity of the trap filling can be improved by adjusting the power of the traps. We used a simple V-profile of trap intensities. To achieve even more uniform filling, we propose fine tuning of individual trap power. In Ref. \onlinecite{bell2016bose}, atom density images were used to correct the intensity profile of a time-averaged ring trap to achieve homogeneous filling of the trap. We plan to implement a similar technique to prepare arrays of atomic ensembles with an even distribution of atoms in the array sites.

V-profiles and tilted profiles were used in measurements in section \ref{section II B} and for protocol C, to achieve a more uniform filling of the arrays or the box potential. In some cases, the sequence of multiplexing was also important: the atoms distributed over the array differently if the optical tweezer traps were multiplexed from left to right or from right to left. For protocol A and B it was best to use a random permutation of the array sites. 

In addition to the uniform filling of the array sites, the variance of atom numbers inside individual traps is also important for applications of atomic-ensemble arrays such as quantum simulation. In Ref. \onlinecite{wang2020preparation} a sub-Poissonian variance of the occupation number of a single trap inside an optical tweezer array was observed. Our measurements show above Poissonian distribution for direct loading into optical tweezers, and close to Poissonian distribution for protocol C, but only for the central part of the array.

The repeatability of loading directly from the dipole trap is worse than the repeatability of protocols A, B and C. This is probably mostly due to the higher initial temperature of the atoms, which also limits the observed maximal number of traps in the array to only 25.
The advantage of direct loading is that the intrinsic distribution of atoms within the array is determined only by the density of the atoms in the dipole trap and the mean velocity of the atoms is zero across the whole array.
This is in contrast to protocols A, B and C, where the uneven distribution of atoms is caused by the expansion determined by the initial velocity distribution. In protocols A and C the mean velocity of atoms at the positions of the traps is in general non-zero, because we are trapping an expanding atomic cloud.
Therefore, protocol A causes oscillations within the trap that can be avoided by stopping the expansion as it is done in protocol B. Protocol C is the most promising in terms of the achieved size of the array. However, for this protocol we only show the distribution of atoms during the expansion, but the atoms will oscillate like in protocol A, if we suddenly stop the expansion of the optical tweezer array. Avoiding inducing the oscillations is possible by slowly stopping the expansion.

This article provides an analysis of time-multiplexed optical tweezers and presents the limitations that the temperature of the atoms sets on the number of array sites.  We conclude that, close to the BEC critical temperature, an array of about a hundred atomic-ensembles could be prepared with the proposed improvements regarding trap intensities. The arrays can be of an arbitrary geometry in a plane, not only in one dimension, if combined with a light-sheet instead of a dimple beam. Furthermore, time-multiplexing can be used in combination with multi-frequency driving or even DMDs or SLMs to achieve hundreds of traps that could be used for quantum simulators.

\begin{acknowledgments}

We thank Rok Žitko for his comments and discussions.
This work was supported by the Slovenian Research Agency (Research Core Fundings No. P1-0125, No. P1-0099 and No. P1-0416).

\end{acknowledgments}

\bibliography{tweezers}

\begin{thebibliography}{32}%
\makeatletter
\providecommand \@ifxundefined [1]{%
 \@ifx{#1\undefined}
}%
\providecommand \@ifnum [1]{%
 \ifnum #1\expandafter \@firstoftwo
 \else \expandafter \@secondoftwo
 \fi
}%
\providecommand \@ifx [1]{%
 \ifx #1\expandafter \@firstoftwo
 \else \expandafter \@secondoftwo
 \fi
}%
\providecommand \natexlab [1]{#1}%
\providecommand \enquote  [1]{``#1''}%
\providecommand \bibnamefont  [1]{#1}%
\providecommand \bibfnamefont [1]{#1}%
\providecommand \citenamefont [1]{#1}%
\providecommand \href@noop [0]{\@secondoftwo}%
\providecommand \href [0]{\begingroup \@sanitize@url \@href}%
\providecommand \@href[1]{\@@startlink{#1}\@@href}%
\providecommand \@@href[1]{\endgroup#1\@@endlink}%
\providecommand \@sanitize@url [0]{\catcode `\\12\catcode `\$12\catcode
  `\&12\catcode `\#12\catcode `\^12\catcode `\_12\catcode `\%12\relax}%
\providecommand \@@startlink[1]{}%
\providecommand \@@endlink[0]{}%
\providecommand \url  [0]{\begingroup\@sanitize@url \@url }%
\providecommand \@url [1]{\endgroup\@href {#1}{\urlprefix }}%
\providecommand \urlprefix  [0]{URL }%
\providecommand \Eprint [0]{\href }%
\providecommand \doibase [0]{https://doi.org/}%
\providecommand \selectlanguage [0]{\@gobble}%
\providecommand \bibinfo  [0]{\@secondoftwo}%
\providecommand \bibfield  [0]{\@secondoftwo}%
\providecommand \translation [1]{[#1]}%
\providecommand \BibitemOpen [0]{}%
\providecommand \bibitemStop [0]{}%
\providecommand \bibitemNoStop [0]{.\EOS\space}%
\providecommand \EOS [0]{\spacefactor3000\relax}%
\providecommand \BibitemShut  [1]{\csname bibitem#1\endcsname}%
\let\auto@bib@innerbib\@empty
\bibitem [{\citenamefont {Wang}\ \emph {et~al.}(2020)\citenamefont {Wang},
  \citenamefont {Shevate}, \citenamefont {Wintermantel}, \citenamefont
  {Morgado}, \citenamefont {Lochead},\ and\ \citenamefont
  {Whitlock}}]{wang2020preparation}%
  \BibitemOpen
  \bibfield  {author} {\bibinfo {author} {\bibfnamefont {Y.}~\bibnamefont
  {Wang}}, \bibinfo {author} {\bibfnamefont {S.}~\bibnamefont {Shevate}},
  \bibinfo {author} {\bibfnamefont {T.~M.}\ \bibnamefont {Wintermantel}},
  \bibinfo {author} {\bibfnamefont {M.}~\bibnamefont {Morgado}}, \bibinfo
  {author} {\bibfnamefont {G.}~\bibnamefont {Lochead}},\ and\ \bibinfo {author}
  {\bibfnamefont {S.}~\bibnamefont {Whitlock}},\ }\bibfield  {title} {\bibinfo
  {title} {Preparation of hundreds of microscopic atomic ensembles in optical
  tweezer arrays},\ }\href@noop {} {\bibfield  {journal} {\bibinfo  {journal}
  {npj Quantum Information}\ }\textbf {\bibinfo {volume} {6}},\ \bibinfo
  {pages} {1} (\bibinfo {year} {2020})}\BibitemShut {NoStop}%
\bibitem [{\citenamefont {Endres}\ \emph {et~al.}(2016)\citenamefont {Endres},
  \citenamefont {Bernien}, \citenamefont {Keesling}, \citenamefont {Levine},
  \citenamefont {Anschuetz}, \citenamefont {Krajenbrink}, \citenamefont
  {Senko}, \citenamefont {Vuleti{\'{c}}}, \citenamefont {Greiner},\ and\
  \citenamefont {Lukin}}]{endres2016atom}%
  \BibitemOpen
  \bibfield  {author} {\bibinfo {author} {\bibfnamefont {M.}~\bibnamefont
  {Endres}}, \bibinfo {author} {\bibfnamefont {H.}~\bibnamefont {Bernien}},
  \bibinfo {author} {\bibfnamefont {A.}~\bibnamefont {Keesling}}, \bibinfo
  {author} {\bibfnamefont {H.}~\bibnamefont {Levine}}, \bibinfo {author}
  {\bibfnamefont {E.~R.}\ \bibnamefont {Anschuetz}}, \bibinfo {author}
  {\bibfnamefont {A.}~\bibnamefont {Krajenbrink}}, \bibinfo {author}
  {\bibfnamefont {C.}~\bibnamefont {Senko}}, \bibinfo {author} {\bibfnamefont
  {V.}~\bibnamefont {Vuleti{\'{c}}}}, \bibinfo {author} {\bibfnamefont
  {M.}~\bibnamefont {Greiner}},\ and\ \bibinfo {author} {\bibfnamefont {M.~D.}\
  \bibnamefont {Lukin}},\ }\bibfield  {title} {\bibinfo {title} {Atom-by-atom
  assembly of defect-free one-dimensional cold atom arrays},\ }\href@noop {}
  {\bibfield  {journal} {\bibinfo  {journal} {Science}\ }\textbf {\bibinfo
  {volume} {354}},\ \bibinfo {pages} {1024} (\bibinfo {year}
  {2016})}\BibitemShut {NoStop}%
\bibitem [{\citenamefont {Bernien}\ \emph {et~al.}(2017)\citenamefont
  {Bernien}, \citenamefont {Schwartz}, \citenamefont {Keesling}, \citenamefont
  {Levine}, \citenamefont {Omran}, \citenamefont {Pichler}, \citenamefont
  {Choi}, \citenamefont {Zibrov}, \citenamefont {Endres}, \citenamefont
  {Greiner}, \citenamefont {Vuleti{\'{c}}},\ and\ \citenamefont
  {Lukin}}]{bernien2017probing}%
  \BibitemOpen
  \bibfield  {author} {\bibinfo {author} {\bibfnamefont {H.}~\bibnamefont
  {Bernien}}, \bibinfo {author} {\bibfnamefont {S.}~\bibnamefont {Schwartz}},
  \bibinfo {author} {\bibfnamefont {A.}~\bibnamefont {Keesling}}, \bibinfo
  {author} {\bibfnamefont {H.}~\bibnamefont {Levine}}, \bibinfo {author}
  {\bibfnamefont {A.}~\bibnamefont {Omran}}, \bibinfo {author} {\bibfnamefont
  {H.}~\bibnamefont {Pichler}}, \bibinfo {author} {\bibfnamefont
  {S.}~\bibnamefont {Choi}}, \bibinfo {author} {\bibfnamefont {A.~S.}\
  \bibnamefont {Zibrov}}, \bibinfo {author} {\bibfnamefont {M.}~\bibnamefont
  {Endres}}, \bibinfo {author} {\bibfnamefont {M.}~\bibnamefont {Greiner}},
  \bibinfo {author} {\bibfnamefont {V.}~\bibnamefont {Vuleti{\'{c}}}},\ and\
  \bibinfo {author} {\bibfnamefont {M.~D.}\ \bibnamefont {Lukin}},\ }\bibfield
  {title} {\bibinfo {title} {Probing many-body dynamics on a 51-atom quantum
  simulator},\ }\href@noop {} {\bibfield  {journal} {\bibinfo  {journal}
  {Nature}\ }\textbf {\bibinfo {volume} {551}},\ \bibinfo {pages} {579}
  (\bibinfo {year} {2017})}\BibitemShut {NoStop}%
\bibitem [{\citenamefont {Pu}\ \emph {et~al.}(2018)\citenamefont {Pu},
  \citenamefont {Wu}, \citenamefont {Jiang}, \citenamefont {Chang},
  \citenamefont {Li}, \citenamefont {Zhang},\ and\ \citenamefont
  {Duan}}]{pu2018experimental}%
  \BibitemOpen
  \bibfield  {author} {\bibinfo {author} {\bibfnamefont {Y.}~\bibnamefont
  {Pu}}, \bibinfo {author} {\bibfnamefont {Y.}~\bibnamefont {Wu}}, \bibinfo
  {author} {\bibfnamefont {N.}~\bibnamefont {Jiang}}, \bibinfo {author}
  {\bibfnamefont {W.}~\bibnamefont {Chang}}, \bibinfo {author} {\bibfnamefont
  {C.}~\bibnamefont {Li}}, \bibinfo {author} {\bibfnamefont {S.}~\bibnamefont
  {Zhang}},\ and\ \bibinfo {author} {\bibfnamefont {L.}~\bibnamefont {Duan}},\
  }\bibfield  {title} {\bibinfo {title} {Experimental entanglement of 25
  individually accessible atomic quantum interfaces},\ }\href@noop {}
  {\bibfield  {journal} {\bibinfo  {journal} {Science advances}\ }\textbf
  {\bibinfo {volume} {4}},\ \bibinfo {pages} {eaar3931} (\bibinfo {year}
  {2018})}\BibitemShut {NoStop}%
\bibitem [{\citenamefont {Carpentier}\ \emph {et~al.}(2008)\citenamefont
  {Carpentier}, \citenamefont {Belmonte-Beitia}, \citenamefont {Michinel},\
  and\ \citenamefont {Rodas-Verde}}]{carpentier2008laser}%
  \BibitemOpen
  \bibfield  {author} {\bibinfo {author} {\bibfnamefont {A.~V.}\ \bibnamefont
  {Carpentier}}, \bibinfo {author} {\bibfnamefont {J.}~\bibnamefont
  {Belmonte-Beitia}}, \bibinfo {author} {\bibfnamefont {H.}~\bibnamefont
  {Michinel}},\ and\ \bibinfo {author} {\bibfnamefont {M.~I.}\ \bibnamefont
  {Rodas-Verde}},\ }\bibfield  {title} {\bibinfo {title} {Laser tweezers for
  atomic solitons},\ }\href@noop {} {\bibfield  {journal} {\bibinfo  {journal}
  {Journal of Modern Optics}\ }\textbf {\bibinfo {volume} {55}},\ \bibinfo
  {pages} {2819} (\bibinfo {year} {2008})}\BibitemShut {NoStop}%
\bibitem [{\citenamefont {Beugnon}\ \emph {et~al.}(2007)\citenamefont
  {Beugnon}, \citenamefont {Tuchendler}, \citenamefont {Marion}, \citenamefont
  {Ga{\"e}tan}, \citenamefont {Miroshnychenko}, \citenamefont {Sortais},
  \citenamefont {Lance}, \citenamefont {Jones}, \citenamefont {Messin},
  \citenamefont {Browaeys},\ and\ \citenamefont {Grangier}}]{beugnon2007two}%
  \BibitemOpen
  \bibfield  {author} {\bibinfo {author} {\bibfnamefont {J.}~\bibnamefont
  {Beugnon}}, \bibinfo {author} {\bibfnamefont {C.}~\bibnamefont {Tuchendler}},
  \bibinfo {author} {\bibfnamefont {H.}~\bibnamefont {Marion}}, \bibinfo
  {author} {\bibfnamefont {A.}~\bibnamefont {Ga{\"e}tan}}, \bibinfo {author}
  {\bibfnamefont {Y.}~\bibnamefont {Miroshnychenko}}, \bibinfo {author}
  {\bibfnamefont {Y.~R.}\ \bibnamefont {Sortais}}, \bibinfo {author}
  {\bibfnamefont {A.~M.}\ \bibnamefont {Lance}}, \bibinfo {author}
  {\bibfnamefont {M.~P.}\ \bibnamefont {Jones}}, \bibinfo {author}
  {\bibfnamefont {G.}~\bibnamefont {Messin}}, \bibinfo {author} {\bibfnamefont
  {A.}~\bibnamefont {Browaeys}},\ and\ \bibinfo {author} {\bibfnamefont
  {P.}~\bibnamefont {Grangier}},\ }\bibfield  {title} {\bibinfo {title}
  {Two-dimensional transport and transfer of a single atomic qubit in optical
  tweezers},\ }\href@noop {} {\bibfield  {journal} {\bibinfo  {journal} {Nature
  Physics}\ }\textbf {\bibinfo {volume} {3}},\ \bibinfo {pages} {696} (\bibinfo
  {year} {2007})}\BibitemShut {NoStop}%
\bibitem [{\citenamefont {Gustavson}\ \emph {et~al.}(2001)\citenamefont
  {Gustavson}, \citenamefont {Chikkatur}, \citenamefont {Leanhardt},
  \citenamefont {G{\"o}rlitz}, \citenamefont {Gupta}, \citenamefont
  {Pritchard},\ and\ \citenamefont {Ketterle}}]{gustavson2001transport}%
  \BibitemOpen
  \bibfield  {author} {\bibinfo {author} {\bibfnamefont {T.~L.}\ \bibnamefont
  {Gustavson}}, \bibinfo {author} {\bibfnamefont {A.~P.}\ \bibnamefont
  {Chikkatur}}, \bibinfo {author} {\bibfnamefont {A.~E.}\ \bibnamefont
  {Leanhardt}}, \bibinfo {author} {\bibfnamefont {A.}~\bibnamefont
  {G{\"o}rlitz}}, \bibinfo {author} {\bibfnamefont {S.}~\bibnamefont {Gupta}},
  \bibinfo {author} {\bibfnamefont {D.~E.}\ \bibnamefont {Pritchard}},\ and\
  \bibinfo {author} {\bibfnamefont {W.}~\bibnamefont {Ketterle}},\ }\bibfield
  {title} {\bibinfo {title} {Transport of {Bose}-{Einstein} condensates with
  optical tweezers},\ }\href@noop {} {\bibfield  {journal} {\bibinfo  {journal}
  {Physical Review Letters}\ }\textbf {\bibinfo {volume} {88}},\ \bibinfo
  {pages} {020401} (\bibinfo {year} {2001})}\BibitemShut {NoStop}%
\bibitem [{\citenamefont {Rakonjac}\ \emph {et~al.}(2012)\citenamefont
  {Rakonjac}, \citenamefont {Deb}, \citenamefont {Hoinka}, \citenamefont
  {Hudson}, \citenamefont {Sawyer},\ and\ \citenamefont
  {Kj{\ae}rgaard}}]{rakonjac2012laser}%
  \BibitemOpen
  \bibfield  {author} {\bibinfo {author} {\bibfnamefont {A.}~\bibnamefont
  {Rakonjac}}, \bibinfo {author} {\bibfnamefont {A.~B.}\ \bibnamefont {Deb}},
  \bibinfo {author} {\bibfnamefont {S.}~\bibnamefont {Hoinka}}, \bibinfo
  {author} {\bibfnamefont {D.}~\bibnamefont {Hudson}}, \bibinfo {author}
  {\bibfnamefont {B.~J.}\ \bibnamefont {Sawyer}},\ and\ \bibinfo {author}
  {\bibfnamefont {N.}~\bibnamefont {Kj{\ae}rgaard}},\ }\bibfield  {title}
  {\bibinfo {title} {Laser based accelerator for ultracold atoms},\ }\href@noop
  {} {\bibfield  {journal} {\bibinfo  {journal} {Optics Letters}\ }\textbf
  {\bibinfo {volume} {37}},\ \bibinfo {pages} {1085} (\bibinfo {year}
  {2012})}\BibitemShut {NoStop}%
\bibitem [{\citenamefont {Bell}\ \emph {et~al.}(2016)\citenamefont {Bell},
  \citenamefont {Glidden}, \citenamefont {Humbert}, \citenamefont {Bromley},
  \citenamefont {Haine}, \citenamefont {Davis}, \citenamefont {Neely},
  \citenamefont {Baker},\ and\ \citenamefont
  {Rubinsztein-Dunlop}}]{bell2016bose}%
  \BibitemOpen
  \bibfield  {author} {\bibinfo {author} {\bibfnamefont {T.~A.}\ \bibnamefont
  {Bell}}, \bibinfo {author} {\bibfnamefont {J.~A.~P.}\ \bibnamefont
  {Glidden}}, \bibinfo {author} {\bibfnamefont {L.}~\bibnamefont {Humbert}},
  \bibinfo {author} {\bibfnamefont {M.~W.~J.}\ \bibnamefont {Bromley}},
  \bibinfo {author} {\bibfnamefont {S.~A.}\ \bibnamefont {Haine}}, \bibinfo
  {author} {\bibfnamefont {M.~J.}\ \bibnamefont {Davis}}, \bibinfo {author}
  {\bibfnamefont {T.~W.}\ \bibnamefont {Neely}}, \bibinfo {author}
  {\bibfnamefont {M.~A.}\ \bibnamefont {Baker}},\ and\ \bibinfo {author}
  {\bibfnamefont {H.}~\bibnamefont {Rubinsztein-Dunlop}},\ }\bibfield  {title}
  {\bibinfo {title} {Bose-{Einstein} condensation in large time-averaged
  optical ring potentials},\ }\href@noop {} {\bibfield  {journal} {\bibinfo
  {journal} {New journal of Physics}\ }\textbf {\bibinfo {volume} {18}},\
  \bibinfo {pages} {035003} (\bibinfo {year} {2016})}\BibitemShut {NoStop}%
\bibitem [{\citenamefont {Boyer}\ \emph {et~al.}(2006)\citenamefont {Boyer},
  \citenamefont {Godun}, \citenamefont {Smirne}, \citenamefont {Cassettari},
  \citenamefont {Chandrashekar}, \citenamefont {Deb}, \citenamefont {Laczik},\
  and\ \citenamefont {Foot}}]{boyer2006dynamic}%
  \BibitemOpen
  \bibfield  {author} {\bibinfo {author} {\bibfnamefont {V.}~\bibnamefont
  {Boyer}}, \bibinfo {author} {\bibfnamefont {R.~M.}\ \bibnamefont {Godun}},
  \bibinfo {author} {\bibfnamefont {G.}~\bibnamefont {Smirne}}, \bibinfo
  {author} {\bibfnamefont {D.}~\bibnamefont {Cassettari}}, \bibinfo {author}
  {\bibfnamefont {C.~M.}\ \bibnamefont {Chandrashekar}}, \bibinfo {author}
  {\bibfnamefont {A.~B.}\ \bibnamefont {Deb}}, \bibinfo {author} {\bibfnamefont
  {Z.~J.}\ \bibnamefont {Laczik}},\ and\ \bibinfo {author} {\bibfnamefont
  {C.~J.}\ \bibnamefont {Foot}},\ }\bibfield  {title} {\bibinfo {title}
  {Dynamic manipulation of {Bose}-{Einstein} condensates with a spatial light
  modulator},\ }\href@noop {} {\bibfield  {journal} {\bibinfo  {journal}
  {Physical Review A}\ }\textbf {\bibinfo {volume} {73}},\ \bibinfo {pages}
  {031402(R)} (\bibinfo {year} {2006})}\BibitemShut {NoStop}%
\bibitem [{\citenamefont {Muldoon}\ \emph {et~al.}(2012)\citenamefont
  {Muldoon}, \citenamefont {Brandt}, \citenamefont {Dong}, \citenamefont
  {Stuart}, \citenamefont {Brainis}, \citenamefont {Himsworth},\ and\
  \citenamefont {Kuhn}}]{muldoon2012control}%
  \BibitemOpen
  \bibfield  {author} {\bibinfo {author} {\bibfnamefont {C.}~\bibnamefont
  {Muldoon}}, \bibinfo {author} {\bibfnamefont {L.}~\bibnamefont {Brandt}},
  \bibinfo {author} {\bibfnamefont {J.}~\bibnamefont {Dong}}, \bibinfo {author}
  {\bibfnamefont {D.}~\bibnamefont {Stuart}}, \bibinfo {author} {\bibfnamefont
  {E.}~\bibnamefont {Brainis}}, \bibinfo {author} {\bibfnamefont
  {M.}~\bibnamefont {Himsworth}},\ and\ \bibinfo {author} {\bibfnamefont
  {A.}~\bibnamefont {Kuhn}},\ }\bibfield  {title} {\bibinfo {title} {Control
  and manipulation of cold atoms in optical tweezers},\ }\href@noop {}
  {\bibfield  {journal} {\bibinfo  {journal} {New Journal of Physics}\ }\textbf
  {\bibinfo {volume} {14}},\ \bibinfo {pages} {073051} (\bibinfo {year}
  {2012})}\BibitemShut {NoStop}%
\bibitem [{\citenamefont {Gauthier}\ \emph {et~al.}(2016)\citenamefont
  {Gauthier}, \citenamefont {Lenton}, \citenamefont {Parry}, \citenamefont
  {Baker}, \citenamefont {Davis}, \citenamefont {Rubinsztein-Dunlop},\ and\
  \citenamefont {Neely}}]{gauthier2016direct}%
  \BibitemOpen
  \bibfield  {author} {\bibinfo {author} {\bibfnamefont {G.}~\bibnamefont
  {Gauthier}}, \bibinfo {author} {\bibfnamefont {I.}~\bibnamefont {Lenton}},
  \bibinfo {author} {\bibfnamefont {N.~M.}\ \bibnamefont {Parry}}, \bibinfo
  {author} {\bibfnamefont {M.}~\bibnamefont {Baker}}, \bibinfo {author}
  {\bibfnamefont {M.~J.}\ \bibnamefont {Davis}}, \bibinfo {author}
  {\bibfnamefont {H.}~\bibnamefont {Rubinsztein-Dunlop}},\ and\ \bibinfo
  {author} {\bibfnamefont {T.~W.}\ \bibnamefont {Neely}},\ }\bibfield  {title}
  {\bibinfo {title} {Direct imaging of a digital-micromirror device for
  configurable microscopic optical potentials},\ }\href@noop {} {\bibfield
  {journal} {\bibinfo  {journal} {Optica}\ }\textbf {\bibinfo {volume} {3}},\
  \bibinfo {pages} {1136} (\bibinfo {year} {2016})}\BibitemShut {NoStop}%
\bibitem [{\citenamefont {Stuart}\ and\ \citenamefont
  {Kuhn}(2018)}]{stuart2018single}%
  \BibitemOpen
  \bibfield  {author} {\bibinfo {author} {\bibfnamefont {D.}~\bibnamefont
  {Stuart}}\ and\ \bibinfo {author} {\bibfnamefont {A.}~\bibnamefont {Kuhn}},\
  }\bibfield  {title} {\bibinfo {title} {Single-atom trapping and transport in
  {DMD}-controlled optical tweezers},\ }\href@noop {} {\bibfield  {journal}
  {\bibinfo  {journal} {New Journal of Physics}\ }\textbf {\bibinfo {volume}
  {20}},\ \bibinfo {pages} {023013} (\bibinfo {year} {2018})}\BibitemShut
  {NoStop}%
\bibitem [{\citenamefont {Shin}\ \emph {et~al.}(2004)\citenamefont {Shin},
  \citenamefont {Saba}, \citenamefont {Pasquini}, \citenamefont {Ketterle},
  \citenamefont {Pritchard},\ and\ \citenamefont {Leanhardt}}]{shin2004atom}%
  \BibitemOpen
  \bibfield  {author} {\bibinfo {author} {\bibfnamefont {Y.}~\bibnamefont
  {Shin}}, \bibinfo {author} {\bibfnamefont {M.}~\bibnamefont {Saba}}, \bibinfo
  {author} {\bibfnamefont {T.~A.}\ \bibnamefont {Pasquini}}, \bibinfo {author}
  {\bibfnamefont {W.}~\bibnamefont {Ketterle}}, \bibinfo {author}
  {\bibfnamefont {D.~E.}\ \bibnamefont {Pritchard}},\ and\ \bibinfo {author}
  {\bibfnamefont {A.~E.}\ \bibnamefont {Leanhardt}},\ }\bibfield  {title}
  {\bibinfo {title} {Atom interferometry with {Bose}-{Einstein} condensates in
  a double-well potential},\ }\href@noop {} {\bibfield  {journal} {\bibinfo
  {journal} {Physical Review Letters}\ }\textbf {\bibinfo {volume} {92}},\
  \bibinfo {pages} {050405} (\bibinfo {year} {2004})}\BibitemShut {NoStop}%
\bibitem [{\citenamefont {Trypogeorgos}\ \emph {et~al.}(2013)\citenamefont
  {Trypogeorgos}, \citenamefont {Harte}, \citenamefont {Bonnin},\ and\
  \citenamefont {Foot}}]{trypogeorgos2013precise}%
  \BibitemOpen
  \bibfield  {author} {\bibinfo {author} {\bibfnamefont {D.}~\bibnamefont
  {Trypogeorgos}}, \bibinfo {author} {\bibfnamefont {T.}~\bibnamefont {Harte}},
  \bibinfo {author} {\bibfnamefont {A.}~\bibnamefont {Bonnin}},\ and\ \bibinfo
  {author} {\bibfnamefont {C.}~\bibnamefont {Foot}},\ }\bibfield  {title}
  {\bibinfo {title} {Precise shaping of laser light by an acousto-optic
  deflector},\ }\href@noop {} {\bibfield  {journal} {\bibinfo  {journal}
  {Optics Express}\ }\textbf {\bibinfo {volume} {21}},\ \bibinfo {pages}
  {24837} (\bibinfo {year} {2013})}\BibitemShut {NoStop}%
\bibitem [{\citenamefont {Schnelle}\ \emph {et~al.}(2008)\citenamefont
  {Schnelle}, \citenamefont {Van~Ooijen}, \citenamefont {Davis}, \citenamefont
  {Heckenberg},\ and\ \citenamefont
  {Rubinsztein-Dunlop}}]{schnelle2008versatile}%
  \BibitemOpen
  \bibfield  {author} {\bibinfo {author} {\bibfnamefont {S.~K.}\ \bibnamefont
  {Schnelle}}, \bibinfo {author} {\bibfnamefont {E.~D.}\ \bibnamefont
  {Van~Ooijen}}, \bibinfo {author} {\bibfnamefont {M.~J.}\ \bibnamefont
  {Davis}}, \bibinfo {author} {\bibfnamefont {N.~R.}\ \bibnamefont
  {Heckenberg}},\ and\ \bibinfo {author} {\bibfnamefont {H.}~\bibnamefont
  {Rubinsztein-Dunlop}},\ }\bibfield  {title} {\bibinfo {title} {Versatile
  two-dimensional potentials for ultra-cold atoms},\ }\href@noop {} {\bibfield
  {journal} {\bibinfo  {journal} {Optics Express}\ }\textbf {\bibinfo {volume}
  {16}},\ \bibinfo {pages} {1405} (\bibinfo {year} {2008})}\BibitemShut
  {NoStop}%
\bibitem [{\citenamefont {Chisholm}\ \emph {et~al.}(2018)\citenamefont
  {Chisholm}, \citenamefont {Thomas}, \citenamefont {Deb},\ and\ \citenamefont
  {Kj{\ae}rgaard}}]{chisholm2018three}%
  \BibitemOpen
  \bibfield  {author} {\bibinfo {author} {\bibfnamefont {C.~S.}\ \bibnamefont
  {Chisholm}}, \bibinfo {author} {\bibfnamefont {R.}~\bibnamefont {Thomas}},
  \bibinfo {author} {\bibfnamefont {A.~B.}\ \bibnamefont {Deb}},\ and\ \bibinfo
  {author} {\bibfnamefont {N.}~\bibnamefont {Kj{\ae}rgaard}},\ }\bibfield
  {title} {\bibinfo {title} {A three-dimensional steerable optical tweezer
  system for ultracold atoms},\ }\href@noop {} {\bibfield  {journal} {\bibinfo
  {journal} {Review of Scientific Instruments}\ }\textbf {\bibinfo {volume}
  {89}},\ \bibinfo {pages} {103105} (\bibinfo {year} {2018})}\BibitemShut
  {NoStop}%
\bibitem [{\citenamefont {Deb}\ \emph {et~al.}(2014)\citenamefont {Deb},
  \citenamefont {McKellar},\ and\ \citenamefont
  {Kj{\ae}rgaard}}]{deb2014optical}%
  \BibitemOpen
  \bibfield  {author} {\bibinfo {author} {\bibfnamefont {A.~B.}\ \bibnamefont
  {Deb}}, \bibinfo {author} {\bibfnamefont {T.}~\bibnamefont {McKellar}},\ and\
  \bibinfo {author} {\bibfnamefont {N.}~\bibnamefont {Kj{\ae}rgaard}},\
  }\bibfield  {title} {\bibinfo {title} {Optical runaway evaporation for the
  parallel production of multiple {Bose}-{Einstein} condensates},\ }\href@noop
  {} {\bibfield  {journal} {\bibinfo  {journal} {Physical Review A}\ }\textbf
  {\bibinfo {volume} {90}},\ \bibinfo {pages} {051401(R)} (\bibinfo {year}
  {2014})}\BibitemShut {NoStop}%
\bibitem [{\citenamefont {Roberts}\ \emph {et~al.}(2014)\citenamefont
  {Roberts}, \citenamefont {McKellar}, \citenamefont {Fekete}, \citenamefont
  {Rakonjac}, \citenamefont {Deb},\ and\ \citenamefont
  {Kj{\ae}rgaard}}]{roberts2014steerable}%
  \BibitemOpen
  \bibfield  {author} {\bibinfo {author} {\bibfnamefont {K.~O.}\ \bibnamefont
  {Roberts}}, \bibinfo {author} {\bibfnamefont {T.}~\bibnamefont {McKellar}},
  \bibinfo {author} {\bibfnamefont {J.}~\bibnamefont {Fekete}}, \bibinfo
  {author} {\bibfnamefont {A.}~\bibnamefont {Rakonjac}}, \bibinfo {author}
  {\bibfnamefont {A.~B.}\ \bibnamefont {Deb}},\ and\ \bibinfo {author}
  {\bibfnamefont {N.}~\bibnamefont {Kj{\ae}rgaard}},\ }\bibfield  {title}
  {\bibinfo {title} {Steerable optical tweezers for ultracold atom studies},\
  }\href@noop {} {\bibfield  {journal} {\bibinfo  {journal} {Optics letters}\
  }\textbf {\bibinfo {volume} {39}},\ \bibinfo {pages} {2012} (\bibinfo {year}
  {2014})}\BibitemShut {NoStop}%
\bibitem [{\citenamefont {Onofrio}\ \emph {et~al.}(2000)\citenamefont
  {Onofrio}, \citenamefont {Durfee}, \citenamefont {Raman}, \citenamefont
  {K{\"o}hl}, \citenamefont {Kuklewicz},\ and\ \citenamefont
  {Ketterle}}]{onofrio2000surface}%
  \BibitemOpen
  \bibfield  {author} {\bibinfo {author} {\bibfnamefont {R.}~\bibnamefont
  {Onofrio}}, \bibinfo {author} {\bibfnamefont {D.~S.}\ \bibnamefont {Durfee}},
  \bibinfo {author} {\bibfnamefont {C.}~\bibnamefont {Raman}}, \bibinfo
  {author} {\bibfnamefont {M.}~\bibnamefont {K{\"o}hl}}, \bibinfo {author}
  {\bibfnamefont {C.~E.}\ \bibnamefont {Kuklewicz}},\ and\ \bibinfo {author}
  {\bibfnamefont {W.}~\bibnamefont {Ketterle}},\ }\bibfield  {title} {\bibinfo
  {title} {Surface excitations of a {Bose}-{Einstein} condensate},\ }\href@noop
  {} {\bibfield  {journal} {\bibinfo  {journal} {Physical Review Letters}\
  }\textbf {\bibinfo {volume} {84}},\ \bibinfo {pages} {810} (\bibinfo {year}
  {2000})}\BibitemShut {NoStop}%
\bibitem [{\citenamefont {Henderson}\ \emph {et~al.}(2009)\citenamefont
  {Henderson}, \citenamefont {Ryu}, \citenamefont {MacCormick},\ and\
  \citenamefont {Boshier}}]{henderson2009experimental}%
  \BibitemOpen
  \bibfield  {author} {\bibinfo {author} {\bibfnamefont {K.}~\bibnamefont
  {Henderson}}, \bibinfo {author} {\bibfnamefont {C.}~\bibnamefont {Ryu}},
  \bibinfo {author} {\bibfnamefont {C.}~\bibnamefont {MacCormick}},\ and\
  \bibinfo {author} {\bibfnamefont {M.~G.}\ \bibnamefont {Boshier}},\
  }\bibfield  {title} {\bibinfo {title} {Experimental demonstration of painting
  arbitrary and dynamic potentials for {Bose}-{Einstein} condensates},\
  }\href@noop {} {\bibfield  {journal} {\bibinfo  {journal} {New Journal of
  Physics}\ }\textbf {\bibinfo {volume} {11}},\ \bibinfo {pages} {043030}
  (\bibinfo {year} {2009})}\BibitemShut {NoStop}%
\bibitem [{\citenamefont {Zimmermann}\ \emph {et~al.}(2011)\citenamefont
  {Zimmermann}, \citenamefont {Mueller}, \citenamefont {Meineke}, \citenamefont
  {Esslinger},\ and\ \citenamefont {Moritz}}]{zimmermann2011high}%
  \BibitemOpen
  \bibfield  {author} {\bibinfo {author} {\bibfnamefont {B.}~\bibnamefont
  {Zimmermann}}, \bibinfo {author} {\bibfnamefont {T.}~\bibnamefont {Mueller}},
  \bibinfo {author} {\bibfnamefont {J.}~\bibnamefont {Meineke}}, \bibinfo
  {author} {\bibfnamefont {T.}~\bibnamefont {Esslinger}},\ and\ \bibinfo
  {author} {\bibfnamefont {H.}~\bibnamefont {Moritz}},\ }\bibfield  {title}
  {\bibinfo {title} {High-resolution imaging of ultracold fermions in
  microscopically tailored optical potentials},\ }\href@noop {} {\bibfield
  {journal} {\bibinfo  {journal} {New Journal of Physics}\ }\textbf {\bibinfo
  {volume} {13}},\ \bibinfo {pages} {043007} (\bibinfo {year}
  {2011})}\BibitemShut {NoStop}%
\bibitem [{\citenamefont {Wu}\ \emph {et~al.}(2021)\citenamefont {Wu},
  \citenamefont {Liang}, \citenamefont {Tian}, \citenamefont {Yang},
  \citenamefont {Chen}, \citenamefont {Liu}, \citenamefont {Tey},\ and\
  \citenamefont {You}}]{wu2021review}%
  \BibitemOpen
  \bibfield  {author} {\bibinfo {author} {\bibfnamefont {X.}~\bibnamefont
  {Wu}}, \bibinfo {author} {\bibfnamefont {X.}~\bibnamefont {Liang}}, \bibinfo
  {author} {\bibfnamefont {Y.}~\bibnamefont {Tian}}, \bibinfo {author}
  {\bibfnamefont {F.}~\bibnamefont {Yang}}, \bibinfo {author} {\bibfnamefont
  {C.}~\bibnamefont {Chen}}, \bibinfo {author} {\bibfnamefont {Y.-C.}\
  \bibnamefont {Liu}}, \bibinfo {author} {\bibfnamefont {M.~K.}\ \bibnamefont
  {Tey}},\ and\ \bibinfo {author} {\bibfnamefont {L.}~\bibnamefont {You}},\
  }\bibfield  {title} {\bibinfo {title} {A concise review of {Rydberg} atom
  based quantum computation and quantum simulation},\ }\href@noop {} {\bibfield
   {journal} {\bibinfo  {journal} {Chinese Physics B}\ }\textbf {\bibinfo
  {volume} {30}},\ \bibinfo {pages} {020305} (\bibinfo {year}
  {2021})}\BibitemShut {NoStop}%
\bibitem [{\citenamefont {Ebadi}\ \emph {et~al.}(2021)\citenamefont {Ebadi},
  \citenamefont {Wang}, \citenamefont {Levine}, \citenamefont {Keesling},
  \citenamefont {Semeghini}, \citenamefont {Omran}, \citenamefont {Bluvstein},
  \citenamefont {Samajdar}, \citenamefont {Pichler}, \citenamefont {Ho},
  \citenamefont {Choi}, \citenamefont {Sachdev}, \citenamefont {Greiner},
  \citenamefont {Vuleti{\'c}},\ and\ \citenamefont {Lukin}}]{ebadi2021quantum}%
  \BibitemOpen
  \bibfield  {author} {\bibinfo {author} {\bibfnamefont {S.}~\bibnamefont
  {Ebadi}}, \bibinfo {author} {\bibfnamefont {T.~T.}\ \bibnamefont {Wang}},
  \bibinfo {author} {\bibfnamefont {H.}~\bibnamefont {Levine}}, \bibinfo
  {author} {\bibfnamefont {A.}~\bibnamefont {Keesling}}, \bibinfo {author}
  {\bibfnamefont {G.}~\bibnamefont {Semeghini}}, \bibinfo {author}
  {\bibfnamefont {A.}~\bibnamefont {Omran}}, \bibinfo {author} {\bibfnamefont
  {D.}~\bibnamefont {Bluvstein}}, \bibinfo {author} {\bibfnamefont
  {R.}~\bibnamefont {Samajdar}}, \bibinfo {author} {\bibfnamefont
  {H.}~\bibnamefont {Pichler}}, \bibinfo {author} {\bibfnamefont {W.~W.}\
  \bibnamefont {Ho}}, \bibinfo {author} {\bibfnamefont {S.}~\bibnamefont
  {Choi}}, \bibinfo {author} {\bibfnamefont {S.}~\bibnamefont {Sachdev}},
  \bibinfo {author} {\bibfnamefont {M.}~\bibnamefont {Greiner}}, \bibinfo
  {author} {\bibfnamefont {V.}~\bibnamefont {Vuleti{\'c}}},\ and\ \bibinfo
  {author} {\bibfnamefont {M.~D.}\ \bibnamefont {Lukin}},\ }\bibfield  {title}
  {\bibinfo {title} {Quantum phases of matter on a 256-atom programmable
  quantum simulator},\ }\href@noop {} {\bibfield  {journal} {\bibinfo
  {journal} {Nature}\ }\textbf {\bibinfo {volume} {595}},\ \bibinfo {pages}
  {227} (\bibinfo {year} {2021})}\BibitemShut {NoStop}%
\bibitem [{\citenamefont {Ebert}\ \emph {et~al.}(2015)\citenamefont {Ebert},
  \citenamefont {Kwon}, \citenamefont {Walker},\ and\ \citenamefont
  {Saffman}}]{ebert2015coherence}%
  \BibitemOpen
  \bibfield  {author} {\bibinfo {author} {\bibfnamefont {M.}~\bibnamefont
  {Ebert}}, \bibinfo {author} {\bibfnamefont {M.}~\bibnamefont {Kwon}},
  \bibinfo {author} {\bibfnamefont {T.~G.}\ \bibnamefont {Walker}},\ and\
  \bibinfo {author} {\bibfnamefont {M.}~\bibnamefont {Saffman}},\ }\bibfield
  {title} {\bibinfo {title} {Coherence and {Rydberg} blockade of atomic
  ensemble qubits},\ }\href@noop {} {\bibfield  {journal} {\bibinfo  {journal}
  {Physical Review Letters}\ }\textbf {\bibinfo {volume} {115}},\ \bibinfo
  {pages} {093601} (\bibinfo {year} {2015})}\BibitemShut {NoStop}%
\bibitem [{\citenamefont {Spong}\ \emph {et~al.}(2021)\citenamefont {Spong},
  \citenamefont {Jiao}, \citenamefont {Hughes}, \citenamefont {Weatherill},
  \citenamefont {Lesanovsky},\ and\ \citenamefont
  {Adams}}]{spong2021collectively}%
  \BibitemOpen
  \bibfield  {author} {\bibinfo {author} {\bibfnamefont {N.~L.~R.}\
  \bibnamefont {Spong}}, \bibinfo {author} {\bibfnamefont {Y.}~\bibnamefont
  {Jiao}}, \bibinfo {author} {\bibfnamefont {O.~D.~W.}\ \bibnamefont {Hughes}},
  \bibinfo {author} {\bibfnamefont {K.~J.}\ \bibnamefont {Weatherill}},
  \bibinfo {author} {\bibfnamefont {I.}~\bibnamefont {Lesanovsky}},\ and\
  \bibinfo {author} {\bibfnamefont {C.~S.}\ \bibnamefont {Adams}},\ }\bibfield
  {title} {\bibinfo {title} {Collectively encoded {Rydberg} qubit},\
  }\href@noop {} {\bibfield  {journal} {\bibinfo  {journal} {Physical Review
  Letters}\ }\textbf {\bibinfo {volume} {127}},\ \bibinfo {pages} {063604}
  (\bibinfo {year} {2021})}\BibitemShut {NoStop}%
\bibitem [{\citenamefont {Xu}\ \emph {et~al.}(2021)\citenamefont {Xu},
  \citenamefont {Venkatramani}, \citenamefont {Cant{\'u}}, \citenamefont
  {{\v{S}}umarac}, \citenamefont {Kl{\"u}sener}, \citenamefont {Lukin},\ and\
  \citenamefont {Vuleti{\'c}}}]{xu2021fast}%
  \BibitemOpen
  \bibfield  {author} {\bibinfo {author} {\bibfnamefont {W.}~\bibnamefont
  {Xu}}, \bibinfo {author} {\bibfnamefont {A.~V.}\ \bibnamefont
  {Venkatramani}}, \bibinfo {author} {\bibfnamefont {S.~H.}\ \bibnamefont
  {Cant{\'u}}}, \bibinfo {author} {\bibfnamefont {T.}~\bibnamefont
  {{\v{S}}umarac}}, \bibinfo {author} {\bibfnamefont {V.}~\bibnamefont
  {Kl{\"u}sener}}, \bibinfo {author} {\bibfnamefont {M.~D.}\ \bibnamefont
  {Lukin}},\ and\ \bibinfo {author} {\bibfnamefont {V.}~\bibnamefont
  {Vuleti{\'c}}},\ }\bibfield  {title} {\bibinfo {title} {Fast preparation and
  detection of a {Rydberg} qubit using atomic ensembles},\ }\href@noop {}
  {\bibfield  {journal} {\bibinfo  {journal} {Physical Reviev Letters}\
  }\textbf {\bibinfo {volume} {127}},\ \bibinfo {pages} {050501} (\bibinfo
  {year} {2021})}\BibitemShut {NoStop}%
\bibitem [{\citenamefont {Ismailov}\ \emph {et~al.}(2021)\citenamefont
  {Ismailov}, \citenamefont {Baizakov},\ and\ \citenamefont
  {Abdullaev}}]{ismailov2021confinement}%
  \BibitemOpen
  \bibfield  {author} {\bibinfo {author} {\bibfnamefont {K.~K.}\ \bibnamefont
  {Ismailov}}, \bibinfo {author} {\bibfnamefont {B.~B.}\ \bibnamefont
  {Baizakov}},\ and\ \bibinfo {author} {\bibfnamefont {F.~K.}\ \bibnamefont
  {Abdullaev}},\ }\bibfield  {title} {\bibinfo {title} {Confinement of bright
  matter-wave solitons on top of a pedestal-shaped potential},\ }\href@noop {}
  {\bibfield  {journal} {\bibinfo  {journal} {Physics Letters A}\ ,\ \bibinfo
  {pages} {127831}} (\bibinfo {year} {2021})}\BibitemShut {NoStop}%
\bibitem [{\citenamefont {Navon}\ \emph {et~al.}(2021)\citenamefont {Navon},
  \citenamefont {Smith},\ and\ \citenamefont {Hadzibabic}}]{navon2021quantum}%
  \BibitemOpen
  \bibfield  {author} {\bibinfo {author} {\bibfnamefont {N.}~\bibnamefont
  {Navon}}, \bibinfo {author} {\bibfnamefont {R.~P.}\ \bibnamefont {Smith}},\
  and\ \bibinfo {author} {\bibfnamefont {Z.}~\bibnamefont {Hadzibabic}},\
  }\bibfield  {title} {\bibinfo {title} {Quantum gases in optical boxes},\
  }\href@noop {} {\bibfield  {journal} {\bibinfo  {journal} {Nature Physics}\
  }\textbf {\bibinfo {volume} {17}},\ \bibinfo {pages} {1334} (\bibinfo {year}
  {2021})}\BibitemShut {NoStop}%
\bibitem [{\citenamefont {Me{\v{z}}nar{\v{s}}i{\v{c}}}\ \emph
  {et~al.}(2019)\citenamefont {Me{\v{z}}nar{\v{s}}i{\v{c}}}, \citenamefont
  {Arh}, \citenamefont {Brence}, \citenamefont {Pi{\v{s}}ljar}, \citenamefont
  {Gosar}, \citenamefont {Gosar}, \citenamefont {{\v{Z}}itko}, \citenamefont
  {Zupani{\v{c}}},\ and\ \citenamefont {Jegli{\v{c}}}}]{meznarsic2019cesium}%
  \BibitemOpen
  \bibfield  {author} {\bibinfo {author} {\bibfnamefont {T.}~\bibnamefont
  {Me{\v{z}}nar{\v{s}}i{\v{c}}}}, \bibinfo {author} {\bibfnamefont
  {T.}~\bibnamefont {Arh}}, \bibinfo {author} {\bibfnamefont {J.}~\bibnamefont
  {Brence}}, \bibinfo {author} {\bibfnamefont {J.}~\bibnamefont
  {Pi{\v{s}}ljar}}, \bibinfo {author} {\bibfnamefont {K.}~\bibnamefont
  {Gosar}}, \bibinfo {author} {\bibfnamefont {{\v{Z}}.}~\bibnamefont {Gosar}},
  \bibinfo {author} {\bibfnamefont {R.}~\bibnamefont {{\v{Z}}itko}}, \bibinfo
  {author} {\bibfnamefont {E.}~\bibnamefont {Zupani{\v{c}}}},\ and\ \bibinfo
  {author} {\bibfnamefont {P.}~\bibnamefont {Jegli{\v{c}}}},\ }\bibfield
  {title} {\bibinfo {title} {Cesium bright matter-wave solitons and soliton
  trains},\ }\href@noop {} {\bibfield  {journal} {\bibinfo  {journal} {Physical
  Review A}\ }\textbf {\bibinfo {volume} {99}},\ \bibinfo {pages} {033625}
  (\bibinfo {year} {2019})}\BibitemShut {NoStop}%
\bibitem [{\citenamefont {Tuchendler}\ \emph {et~al.}(2008)\citenamefont
  {Tuchendler}, \citenamefont {Lance}, \citenamefont {Browaeys}, \citenamefont
  {Sortais},\ and\ \citenamefont {Grangier}}]{tuchendler2008energy}%
  \BibitemOpen
  \bibfield  {author} {\bibinfo {author} {\bibfnamefont {C.}~\bibnamefont
  {Tuchendler}}, \bibinfo {author} {\bibfnamefont {A.~M.}\ \bibnamefont
  {Lance}}, \bibinfo {author} {\bibfnamefont {A.}~\bibnamefont {Browaeys}},
  \bibinfo {author} {\bibfnamefont {Y.~R.~P.}\ \bibnamefont {Sortais}},\ and\
  \bibinfo {author} {\bibfnamefont {P.}~\bibnamefont {Grangier}},\ }\bibfield
  {title} {\bibinfo {title} {Energy distribution and cooling of a single atom
  in an optical tweezer},\ }\href@noop {} {\bibfield  {journal} {\bibinfo
  {journal} {Physical Review A}\ }\textbf {\bibinfo {volume} {78}},\ \bibinfo
  {pages} {033425} (\bibinfo {year} {2008})}\BibitemShut {NoStop}%
\bibitem [{\citenamefont {He}\ \emph {et~al.}(2011)\citenamefont {He},
  \citenamefont {Yang}, \citenamefont {Zhang},\ and\ \citenamefont
  {Wang}}]{he2011extending}%
  \BibitemOpen
  \bibfield  {author} {\bibinfo {author} {\bibfnamefont {J.}~\bibnamefont
  {He}}, \bibinfo {author} {\bibfnamefont {B.-D.}\ \bibnamefont {Yang}},
  \bibinfo {author} {\bibfnamefont {T.-C.}\ \bibnamefont {Zhang}},\ and\
  \bibinfo {author} {\bibfnamefont {J.-M.}\ \bibnamefont {Wang}},\ }\bibfield
  {title} {\bibinfo {title} {Extending a release-and-recapture scheme to single
  atom optical tweezer for effective temperature evaluation},\ }\href@noop {}
  {\bibfield  {journal} {\bibinfo  {journal} {Chinese Physics B}\ }\textbf
  {\bibinfo {volume} {20}},\ \bibinfo {pages} {073701} (\bibinfo {year}
  {2011})}\BibitemShut {NoStop}%
\end{thebibliography}%

\end{document}